\documentclass[a4paper]{article}

\usepackage[pdftex]{graphicx}
\usepackage[utf8]{inputenc}
\usepackage{amsmath}

\DeclareGraphicsExtensions{.jpeg,.pdf}
\title{Rapid self-organised initiation of {\em ad hoc} sensor networks close 
       above the percolation threshold}

\author{Reinert Korsnes\thanks{Norwegian Defence Research Establishment (FFI),
                                 Postboks 25, NO-2027 Kjeller, Norway}}

\begin{document}
\maketitle


\begin{abstract}
This work shows potentials for rapid self-organisation of sensor networks
where nodes collaborate to relay messages to a common data collecting unit (sink node).
The study problem is, in the sense of graph theory, to find
a shortest path tree spanning a weighted graph.
This is a well-studied problem where for example
Dijkstra's algorithm provides a solution for non-negative
edge weights. 
The present contribution shows by simulation examples that simple modifications 
of known distributed approaches here can provide significant improvements 
in performance. 
Phase transition phenomena,
which are known to take place in networks close to percolation thresholds,
may explain these observations. 
An initial method, which here serves as reference, assumes the sink node starts 
organisation of the network (tree) by transmitting a control
message advertising its availability for its neighbours.
These neighbours then advertise their current cost estimate 
for routing a message to the sink.
A node which in this way receives a message implying an improved route to the sink,
advertises its new finding and remembers which neighbouring node the message came from. 
This activity proceeds until there are no more improvements to advertise to neighbours. 
The result is a tree network for cost effective transmission
of messages to the sink (root).
This distributed approach has potential for simple improvements 
which are of interest when minimisation of storage and communication
of network information are a concern.
Fast organisation of the network takes place when
the number $k$ of connections for each node ({\em degree}) 
is close above its critical value for global network percolation 
and at the same time there is a threshold for the nodes 
to decide to advertise network route updates.
\end{abstract}

\section{Transmission within sensor networks}
This contribution shows simulated examples of simple and rapid organisation 
of large {\em ad hoc} sensor networks \cite{Akyildiz2002}. 
The work may have direct interest for design and application of sensor networks.
The present practical examples may also impact theoretical development \cite{Dorogovtsev2003}
and have general interest for understanding network organisation outside
the scope of sensor and computer networks.

The present work is about minimisation of transmission power for sending messages 
between nodes in a sensor network.  
This is directly relevant for underwater sensor networks based on acoustic communication.  
Energy consumption for transmission in this case dominates total energy consumption 
and hence battery  life time \cite[]{Stojanovic2006}.
                  
Rapid network organisation with a minimum of control traffic
and low transmission power is a general protection measure
for {\em ad hoc} sensor networks. 
The amount of control traffic and the level of transmission power 
directly affect the probability to discover and 
to map a sensor network from outside.
Minimisation of storage of network information in nodes
(k-local information) and low processing complexity  
is also a general protection measure reducing 
opportunities for malicious attacks.

\section{The contribution of this work}
Assume a connected weighted graph $(G = (V,E),\omega)$ in the sense of graph theory.
It is a well-studied problem to find the shortest path tree spanning 
such a graph \cite{Wu2004,Epstein1999,Dijkstra1959,Elkin2005,Awerbuch1989}.
Vertices ($V$) are below called 'nodes', edges ($E$) are called 'links' and
weights $\omega$ are called 'link cost'.
This wording is due to the present focus on sensor networks.
The present simulations demonstrate potentials for simple, fast, 
silent (few messages between nodes) and self-organised
buildup of a spanning tree for efficient data relaying from 
any node (sensor) and to a common sink node (root). 
A point is also to minimise storage of network information in single nodes.
Each node has (or can generate) an estimate of a cost  
(a real number) for transmitting
a message back to a node from which it directly receives a message  
(for example a function of the ratio between received and transmitted energy
assuming  the node can measure the received effect and knows the initial output effect). 

The present simulations are for 4000 nodes with random positions
within a flat (2d) square area of $4000 \times 4000$\,$\mbox{m}^{2}$. 
The nodes can transmit messages to each other within a restricted range
(assuming electromagnetic or acoustic communication).
Each node $X$ initially stores (for formal reasons) a cost estimate equal 
to infinity ($C(X,S) = \infty$) for relaying a message to the sink node $S$.
The sink node at first advertises to its neighbours a cost estimate equal zero
for sending a messages to the sink (i.e.\ to itself).
This gives cost estimates for the neighbours and which they further advertise
to their neighbours.
A node $X$, which in general receives a cost estimate form a neighbour $N$ 
(which might be the sink), adds the received cost estimate
from the sender plus a cost to transmit directly back to it. 
This gives a cost estimate to send a message to the sink.
If this estimate represents an improved path to the sink (i.e.\ $C(X,S) < C(X,N) + C(N,S)$),
it updates its cost estimate $C(X,S)$ (setting its value to $C(X,N) + C(N,S)$). 
It then also sets a pointer $E$ to point to the neighbour $N$ (or $E = (X,N)$).
In addition it advertises to its neighbours its latest improved cost estimate.
This procedure proceeds until convergence, 
and the set of such pointers gives a shortest path tree spanning all nodes.

The above method to find a covering minimum path length tree is well-known.
However, 
this work tests out the idea to do the following modifications to this approach: 
\begin{itemize}
\item Restrict the set of edges (neighbours) 
      in the original network (subtract) to the nearest neighbours 
      so that the network is close to loose its global connectivity.   
      This means, in other words, that nodes ignore their most distant neighbours.
\item Restrict advertisement of improved cost estimates 
      (so that nodes report to their neighbours only ''significant'' improvements 
      of their costs estimates for the path to the sink).
\end{itemize}
These restrictions can give a surprisingly fast network layout  
where each node needs to transmit to its neighbours 
typically only 2-4 messages during buildup.

The actual cost for transmitting data here increases with distance
faster than linearly.
This favours relying data through many small hops as compared 
to few large (spatially long) hops.
Hence reduction of actual connections to the 'nearest' neighbours
(in the sense of transmission cost),  allows 
for reduction of radio transmission range and probability 
for communication interference (data packet loss). 

Several authors have discussed connectivity in wireless
{\em ad hoc} networks as a function of number of node neighbours 
\cite{Kleinrock1978,Takagi1984,Xue2004}.
At least eight neighbours for each node are enough to assure connectivity
within a set of nodes with random distribution
on a flat (two-dimensional) area.

Authors have applied ideas on existence of critical transmission power and
also percolation theory to investigate connectivity in {\em ad hoc} sensor networks 
\cite{Meguerdichian2001,Flaxman2004,Ding2008,Krishnamachari2001,Gupta1998,Sanchez1999}.
Flaxman {\em et.\ al} \cite{Flaxman2004} 
had a setting of unreliable nodes and they found
a critical radius to guarantee multi-hop communication
links where nodes have a random distribution
within a square area.
Muthukrishnan and Gopal Pandurangan \cite[]{MuthukrishnanP05}
similarly found critical values for transmission range 
to ensure connectivity and path length estimates
using random graph analysis. 
An intuition behind the present work is to see if network formation 
can benefit from (critical) phenomena, such as change in correlation lengths
or 'temporal alignments' or order,
taking place in networks close to their percolation threshold. 
Tuning of number of neighbours and restrictions for sending control messages 
affect correlation lengths emerging in systems close to their 
phase transitions [see for example Refs.\ \cite{Stainley1971,Goldenfeld1992,Kadanoff2000}].

A reason to pay attention to the present approach is its simplicity.
Alternatives can for example be 
to use dominating (sub-)sets of nodes to reduce
control traffic and data packet collision in {\em ad hoc}
sensor networks 
\cite[]{Wu2008,Shang2008,Thai2007,YLi2005}.
A selected subset of nodes here function as a traffic backbone.
The present work assumes no such structure and dependencies 
in the network adding complexity.
The present method for relaying data to a sink node,  
also requires no network wide (global) 
identity for nodes (it assumes for example no IP addresses).

\section{The sink direction protocol}

\subsection{Finding minimum cost routes to a sink}
\label{se:minimum_cost_route}
This section addresses locally based organisation of
cost optimal multi-hop 
routing from a set of sensor nodes and to a common sink or receiver. 
The approach in this section serves as a reference below and it 
is similar to common search for optimal paths on road maps  
[cf Dijkstra's algorithm \cite{Dijkstra1959}].
\mbox{Eq.\ \ref{eq:acoustic_cost}} here defines the cost for transmitting
a signal between nodes. 
It exemplifies a general class
of cost functions increasing with distance between transmitter
and receiver.
Each node in the established tree network has a pointer 
telling to which neighbour to direct a message
for further routing to the sink (root).
Actual cost functions are additive so that
the total cost of a route is the sum of the cost
for each step along the route. 

A simple method to create a tree network for data routing,
is that any node receiving information about a decreased 
value for the cost of sending a message to the sink,
transmits to its neighbours a message telling
its identity and its new cost value.
A node which receives such a message, 
updates its pointer and cost estimate if it 
represents a better (more cost effective) way to the sink. 
Each node $X$, in other words, keeps a variable $C(X,S)$ 
({\em route cost}) 
where the value is an estimate for the total {\em cost} for transmitting 
a message to the sink $S$.
This variable is, for formal reasons, 
infinite (i.e.\ $C(X,S) = \infty$) for each node
$X$ which has not received a message.
When a node $X$ receives a message   
from a neighbour $N$ giving an improved cost estimate,
i.e.\ if 
\begin{equation}
\label{eq:test1}
C(X,N) + C(N,S) < C(X,S)
\end{equation}
then it sets its pointer $E(X)$ to the edge leading to $N$ ($E = (X,N)$)
which then is part of a finite cost route from $X$ to the sink $S$. 
Section \ref{se:modified_gossips} gives a modified version 
of this simple sink direction method and which possesses improved convergence.

\begin{figure}[htp]
\begin{center}
\resizebox{0.49\textwidth}{!}{\includegraphics{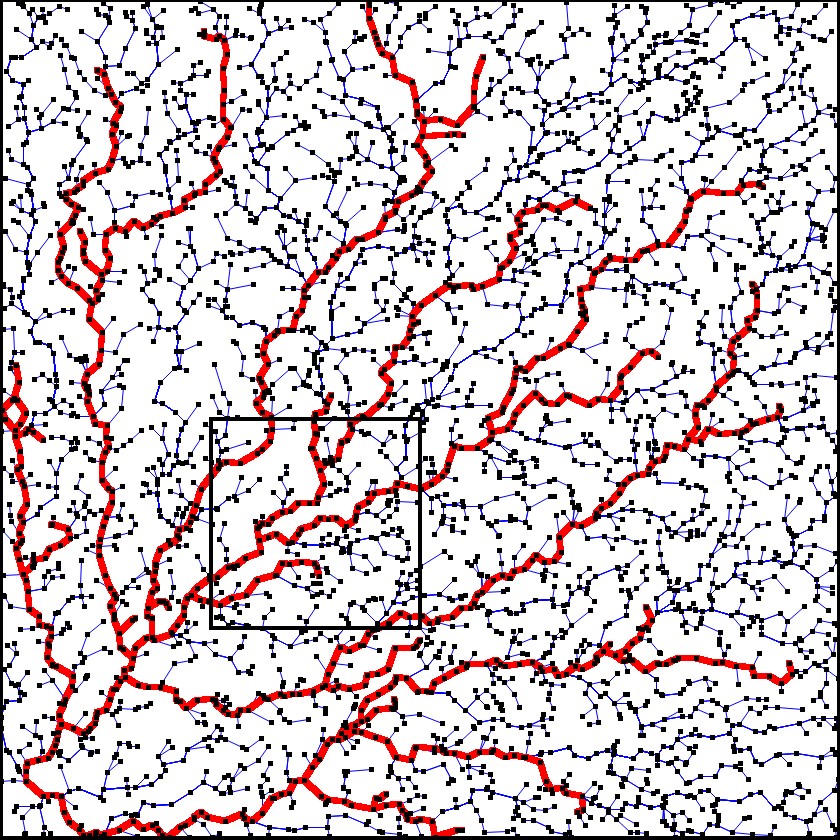}}
\resizebox{0.49\textwidth}{!}{\includegraphics{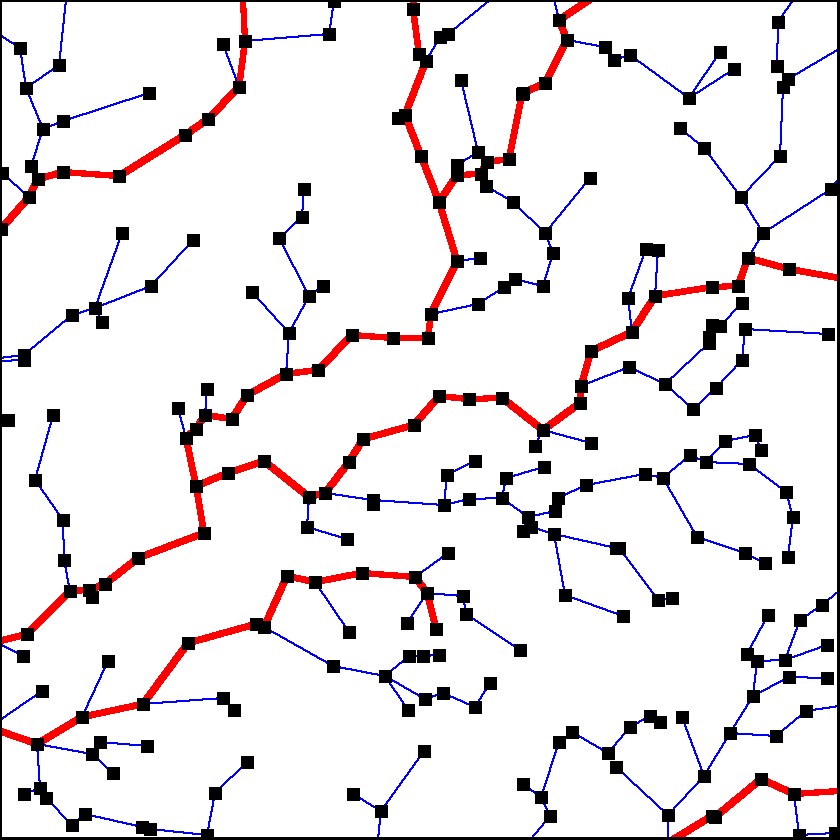}}
\end{center}
\caption{
Left: Example of optimal paths 
through a set of 4000 nodes with random distribution
over an area of $4000 \times 4000$\,$\mbox{m}^2$. 
Transmission range (radius) is 300\,m. 
The sink is close to the origin (with coordinates (200,200) at the lower left corner).
Right: subset (frame) of left image.
Red line shows example of path from node to sink.
These plots results from using the well-known program {\em gnuplot}
and data from the present simulations. 
}
\label{fig:map_cost}
\end{figure}

Figure \ref{fig:map_cost} shows an example from simulation 
of initiation of an energy optimal tree network where transmission cost
per packet, $C$,
scales with range $r$ as proposed by Stojanovic \cite{Stojanovic2006}
for underwater acoustic systems:
\begin{equation}
\label{eq:acoustic_cost}
C \sim (r/a)^{1.5}  \exp(r/b)
\end{equation}
$a$ and $b$ are in this example both for simplicity set to 100\,m.
The simulation is for 4000 nodes with a random distribution
within a square area of $4000 \times 4000$\,$\mbox{m}^2$.
The transmission range is 300\,m.
The average number of neighbours (node degree) is in this case 70
for nodes more than 300\,m from the border.
The nodes can send data packets to their neighbourhood at 
given time steps (for example each second).
Note that the routes tend to consist of small steps
(due to the definition of link cost).

The author made the software for the present simulations 
(which produced data for Figure \ref{fig:map_cost}) 
by direct programming in Ada 2005 (GNU Ada under Linux). 
The size of the actual program illustrates the low complexity 
of the present approach 
(about 300 Ada program statements in total). 
A simulation took less than one minute on a common laptop with 64-bit CPU
(the quickest simulations - below called the ''rapid method'' 
- took less than 4 seconds).
This simulation included generation of random node positions
and link connections as illustrated by Figure \ref{fig:map_cost}.

\section{Modification of the sink direction protocol}
\label{se:modified_gossips}

\subsection{Using link cost for temporal control}
\label{se:link_cost_control}
Assume a network as in Section \ref{se:minimum_cost_route} above.  
Let "agents" start to walk along each edge from a sink node with constant velocity $v$.
Each time an agent arrives at a node, 
it triggers other agents to precede the walk with the same velocity along adjacent paths.
The first arrival at a node in this case gives the shortest path to the sink.
This is an intuitive distributed implementation of 
Dijkstra's breadth-first search  algorithm \cite[]{Dijkstra1959}.
The ''agent walk'' can be defined to be self-avoiding since
repeated walk along the same path is not any optimal path.

Synchronised clocks at the nodes give the opportunity to implement this idea looking
at link cost as ''road length''.
Assume a number of nodes as above.
The sink node, as above, initiates tree network formation by sending 
a message telling its link cost (equal zero). 
Assume the nodes have synchronised clocks
giving the time $t$ elapsed since the sink sent its message starting network organisation.
Any node $X$ (as above) keeps updated its cost estimate $C(X,S)$ 
according to estimates received.
However, it delays to advertise its cost estimate until
the condition $v \cdot t > C(X,S)$ is fulfilled (where $v$ is the assumed ''velocity'' as above
or simply cost per time unit).
Assume here that the probability for simultaneous advertisements is zero. 

Time synchronisation here functions as an alternative to
central synchronisation by the sink node as described in \cite{Awerbuch1989}.
Each node in this case only send one message during the network (tree) formation.
However, access to a common time parameter normally
requires (for example radio) receivers or clocks and 
communication to synchronise them.
The nodes could keep track of time since start of a signal from the sink.
This requires estimates of time for signals in for 
example the water (for underwater sensor networks based on acoustic communication). 
Such time control for sending messages also requires conservative waiting times
(i.e.\ a ''velocity'' $v$ small enough) to assure that a node does not 
receive a better cost estimate after it already has advertised its cost estimate.

Time synchronised search as described above,
seems to fall in the category of centrally designed systems.
I.e.\ the alternative non-synchronised and distributed approach below
may have more general interest for signalling within for example 
biological systems.

\subsection{Application of one-step neighbour lists}
\label{se:nabour_info}

Assume a set of nodes act as in Section \ref{se:minimum_cost_route} above.  
Each node in addition initially collects the identity of the neighbours 
of its neighbours and their links with associated cost (neighbour lists). 
Consider, in this case, a node $X$ receives a message directly from 
a node $Y$ which tells its present cost estimate $C(Y,S)$ 
for the cost to transport a data package to the sink node $S$.   
If $C(X,Y) + C(Y,S) < C(X,S)$), 
then $X$ can (as in Section \ref{se:minimum_cost_route} above)
update its cost estimate $C(X,S)$ and set its pointer $E$ to 
point to $Y$.
It could then advertise the value of its new cost estimate $C(X,S)$.
However, $X$ can use its information about its neighbourhood
to check if it can expect better cost estimates 
from its neighbourhood if there is a multi-hop
route from $Y$ to $X$ via its neighbours giving an even 
better (total) cost estimate for routing data to the sink.
$X$ will in this case wait to transmit its latest 
cost estimate until it receives a better cost estimate
from one of its neighbours.
This will reduce the number of messages transmitted during
network buildup (as compared to the approach of Section \ref{se:minimum_cost_route}).
Note, however, that creation of neighbour lists for each node
requires to send messages reducing the net gain with respect 
to minimising number of messages.

\subsection{Relaxation of condition to inform neighbours on new cost estimates}
\label{se:relaxation}
The sink direction protocol above makes nodes create a pointer system 
converging towards a minimum path tree for transmitting
single messages from a node to the sink.
This work shows by example that it is possible, 
by simple modifications,
significantly to reduce the number of messages during this process.   
Actual modifications are:
\begin{itemize}
\item The following condition (test) 
      replaces \mbox{Eq.\ \ref{eq:test1}}: 
      \begin{equation}
         \label{eq:test1_general}
         C(X,N) + C(N,S) < f \cdot C(X,S)
      \end{equation}
      where the constant $f \ge 1$ defines 
      a threshold for a node to tell neighbours
      about improved cost estimate. 
      The results below are from simulations with 
      $f = 1, 1.01, 1.02, 1.10$.
\item The nodes only listen to their  
       $k$ nearest neighbouring nodes
      (in terms of link cost). 
      The results below are from simulations with 
      $k = 8, 16, 32, 64$.
\end{itemize}

Eq.\ \ref{eq:test1} and \ref{eq:test1_general} define 
the decision of a node $X$ to transmit 
to its neighbours its newest (best) cost estimate $C(X,S)$.
The present simulations employ a random time delay 
from when the node recognises validity of this condition 
(defined by \mbox{Eq.\  \ref{eq:test1}} and \ref{eq:test1_general})
and until the transmission actually takes place. 
This delay time has an uniform distribution 
in the range $1,2,\cdots,100$ time steps (seconds).
Random delay times for sending data packets is a common technique 
to avoid packet collisions.
A node may receive messages from the neighbours during the delay time. 

\begin{figure}[htp]
\begin{center}
\resizebox{0.48\textwidth}{!}{\includegraphics{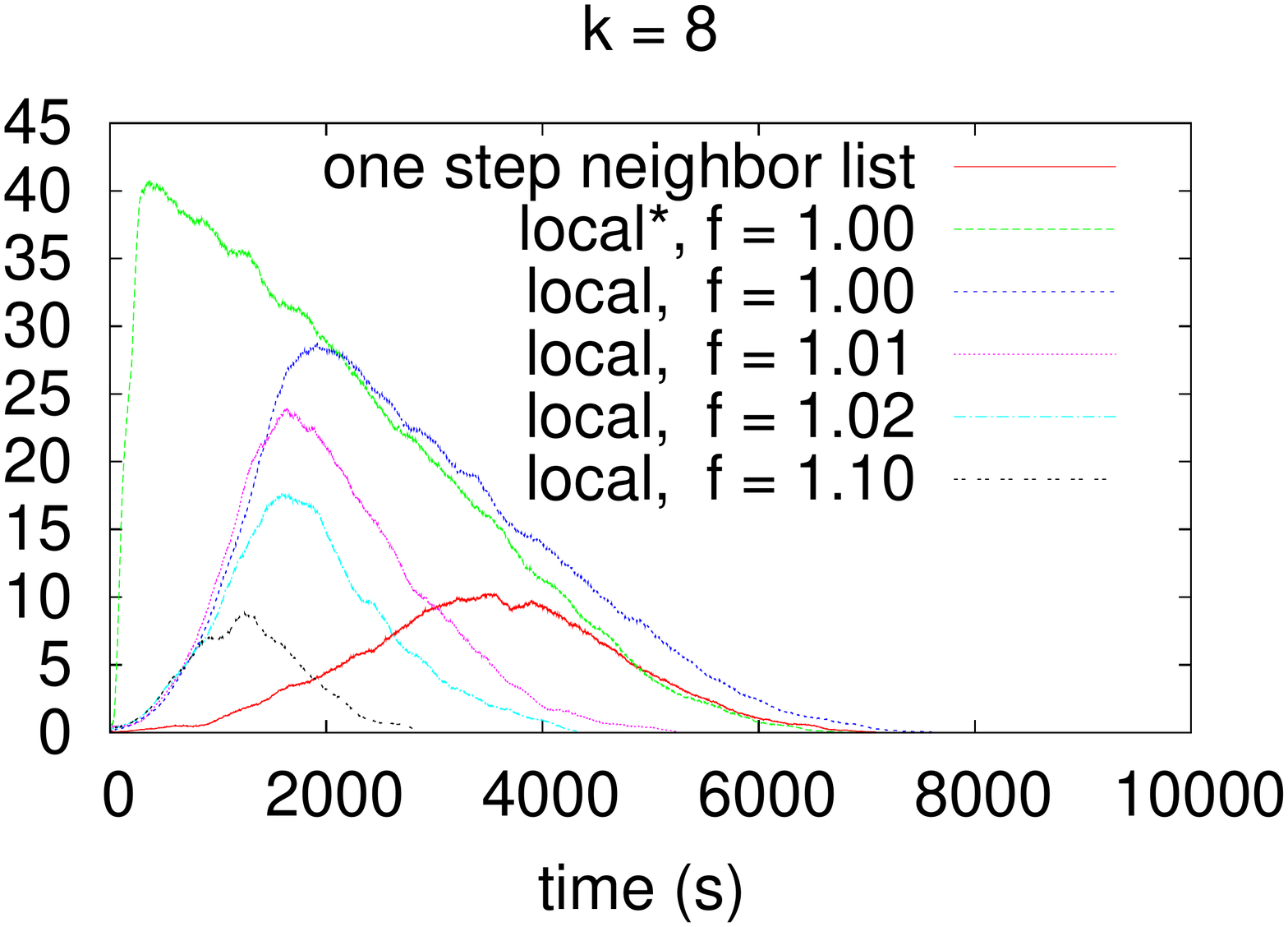}}  
\resizebox{0.48\textwidth}{!}{\includegraphics{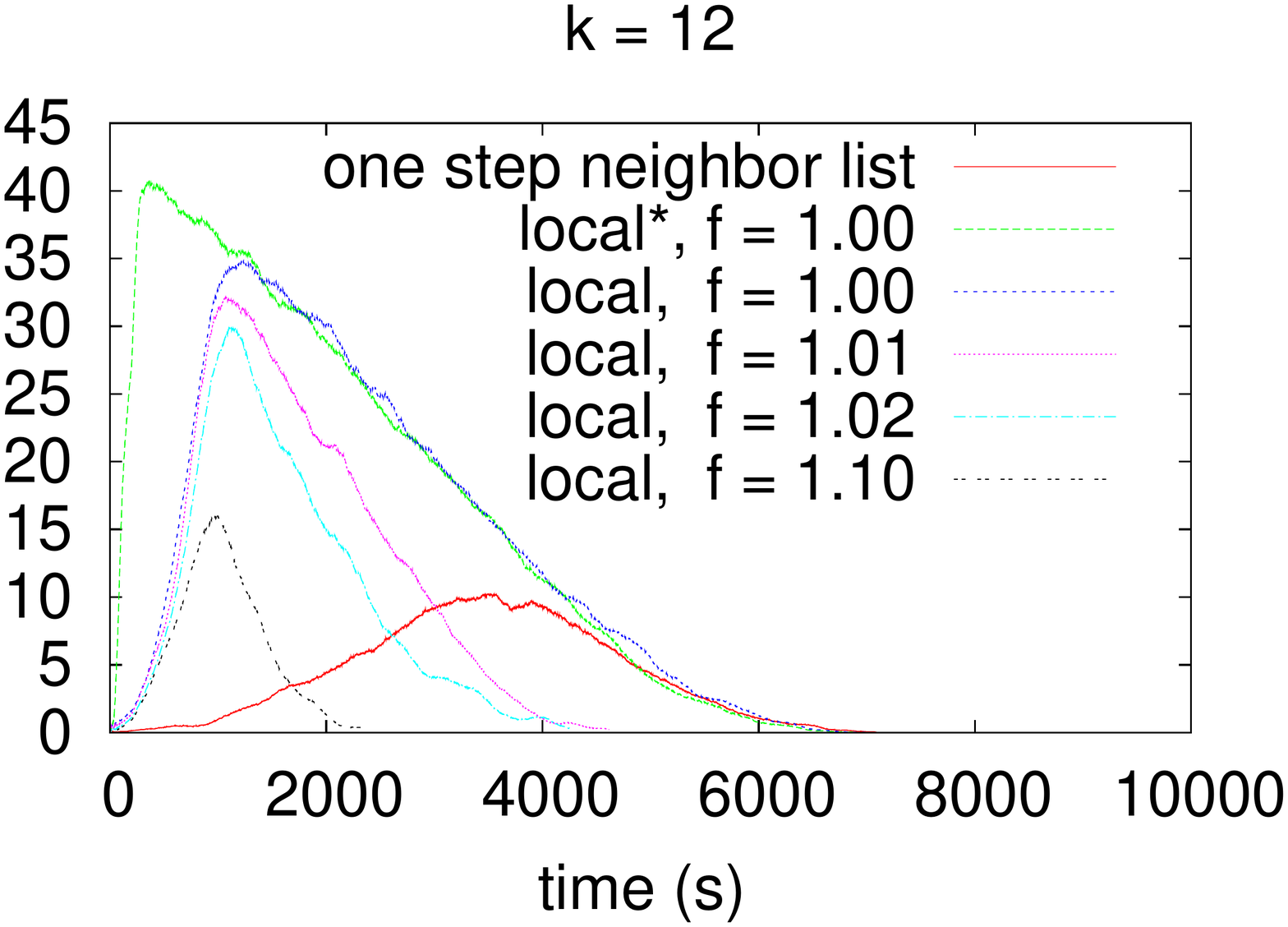}} \\  
\resizebox{0.48\textwidth}{!}{\includegraphics{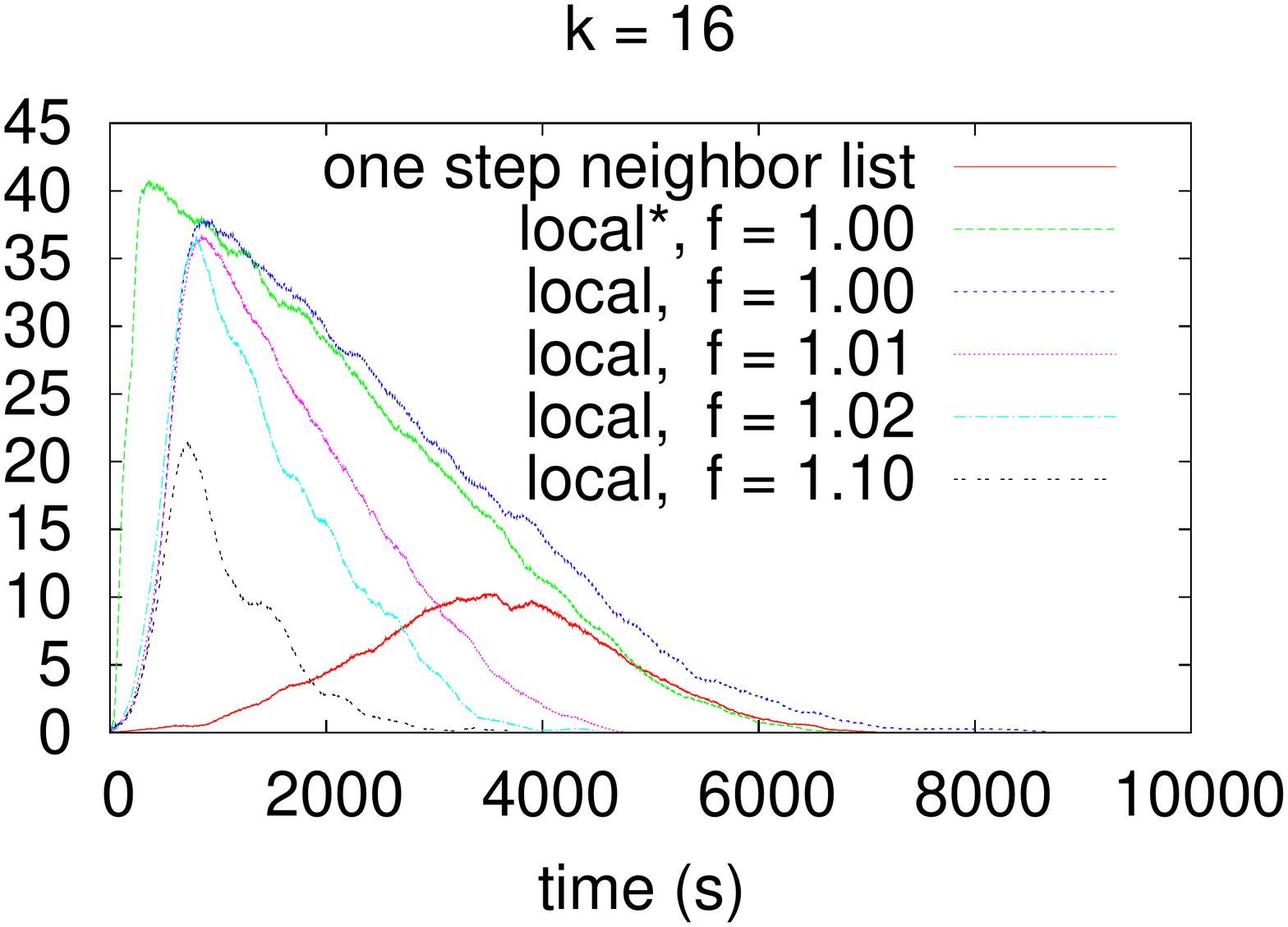}}  
\resizebox{0.48\textwidth}{!}{\includegraphics{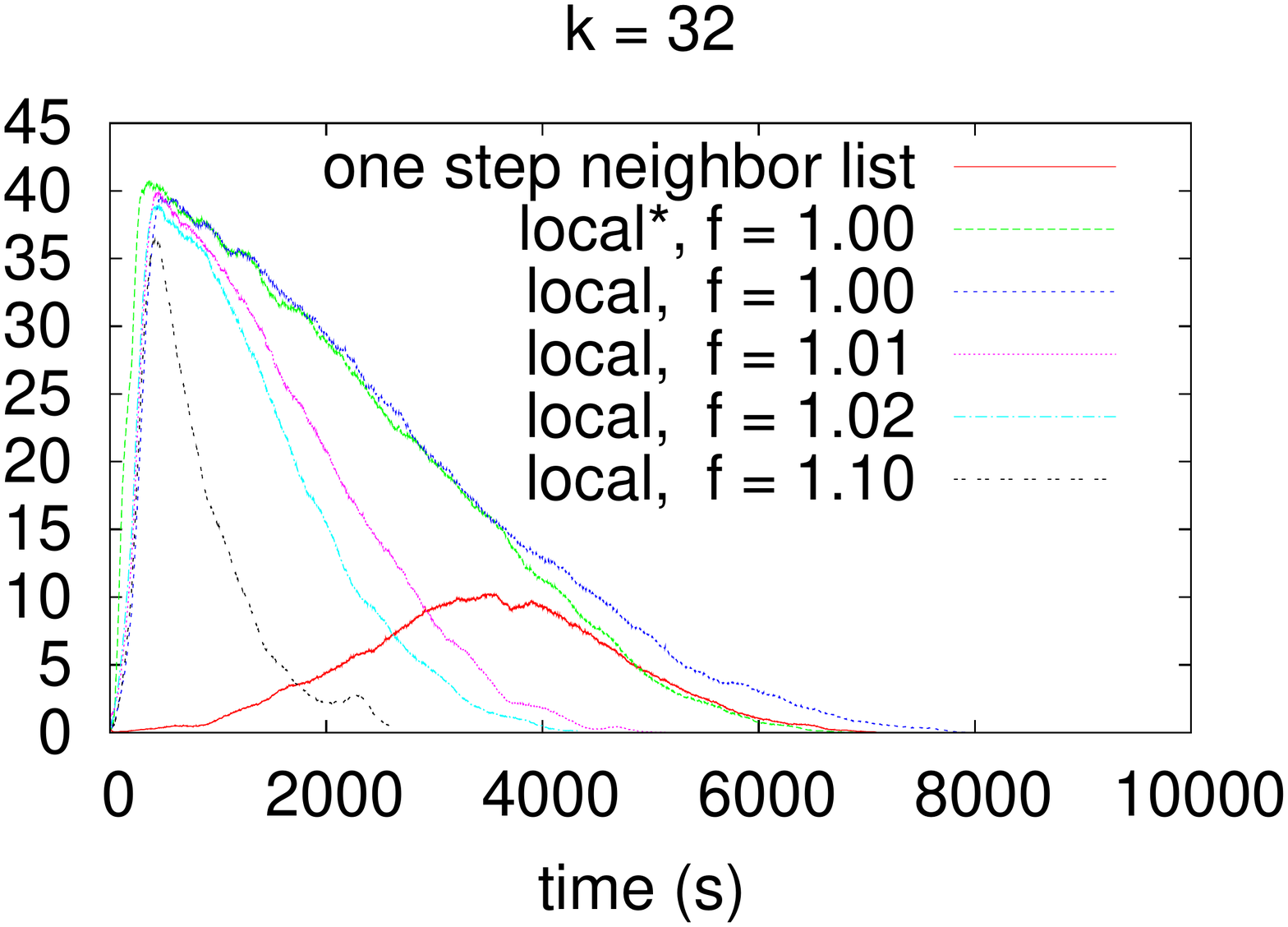}}   
\end{center}
\caption{Time series of number of transmitted data packets during
self-organisation of sensor network.
These results are from the present simulations with communication 
only to the $k$ nearest neighbours where $k = 8, 12, 16, 32$. 
200\,s moving average.
Note the rapid network initiation for $f = 1.1,\, k = 8$.
}
\label{fig:convergence}
\end{figure}

Figure \ref{fig:convergence} shows time series of 
total number of messages (from all 4000 nodes)
produced via the present (simulated) approach.
The upper graph (green, with label "local*, $f = 1$") 
shows results from a naive local procedure where each node
reports any improved cost estimate 
(given by the condition set by \mbox{Eq.\ \ref{eq:test1}}).
The graph next below the upper graph (with label "local, $f = 1$") 
shows cost estimates
where the number $k$ of neighbours are restricted to $k = 8, 12, 16, 32$ and 
$f = 1$ (same as for upper graph).
The red graph shows result from simulation where nodes use neighbourhood 
information as described in Section \ref{se:nabour_info}.

\begin{figure}[htp]
\begin{center}
\resizebox{0.9\textwidth}{!}{\includegraphics{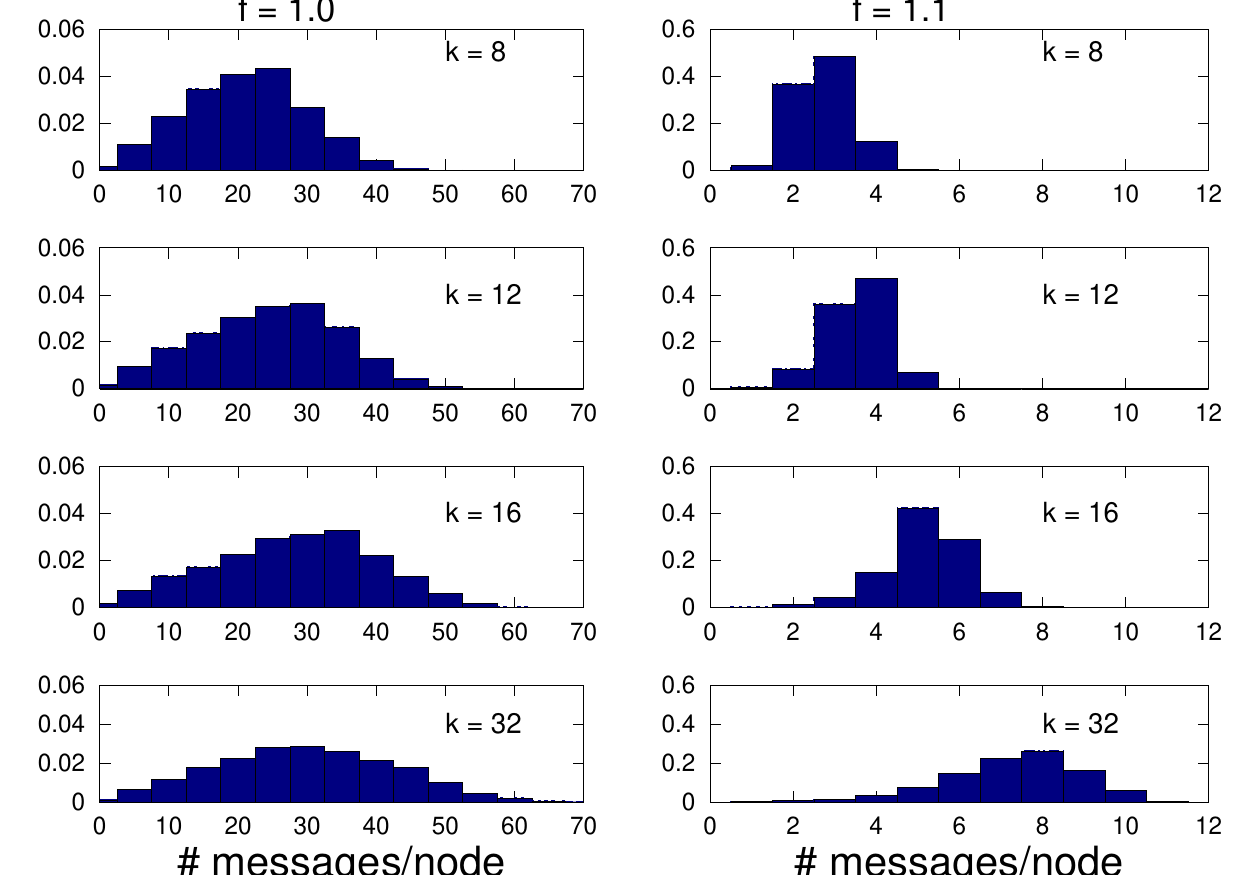}}  
\end{center}
\caption{
         Distribution (histogram) for number of messages sent by
         each of the 4000 nodes in order to generate a tree network 
         to transport of data to a sink node (root). 
         These results are from present simulations with number of neighbours
         $k = 8, 12, 16, 32$.
         Left row of histograms are from simulations with 
         no relaxation on condition to send update information 
         to neighbours (i.e.\ $f = 1$).
         The right row of histograms similarly shows relaxed condition
         $f = 1.1$ (cf Eq.\ \ref{eq:test1_general}) which has
         a significant effect on performance.
}
\label{fig:hist_gossip2a}
\end{figure}

Figure \ref{fig:hist_gossip2a}
shows the distribution (histogram) of number of messages sent by
each node to generate a tree network to relay data to the sink for 
respectively $f = 1$ and $f = 1.1$.  
Note that for the case $f = 1.1$ and 8 neighbours ($k = 8$)
each node needs typically to transmit
only three-four messages to generate the whole tree network.
Hence there is little room for further improvements 
of performance defined as number of calculation and transmissions
of signals (however the resulting tree network is not fully 
cost optimal as noted below).
The number of primitive calculations scales linearly with the number $n$ of nodes.  
Note that the non-local method includes transmissions
to obtain neighbour list from the neighbours. 
The numerical results above do not include this type of initial traffic.

\begin{figure}[htp]
\begin{center}
\resizebox{0.6\textwidth}{!}{\includegraphics{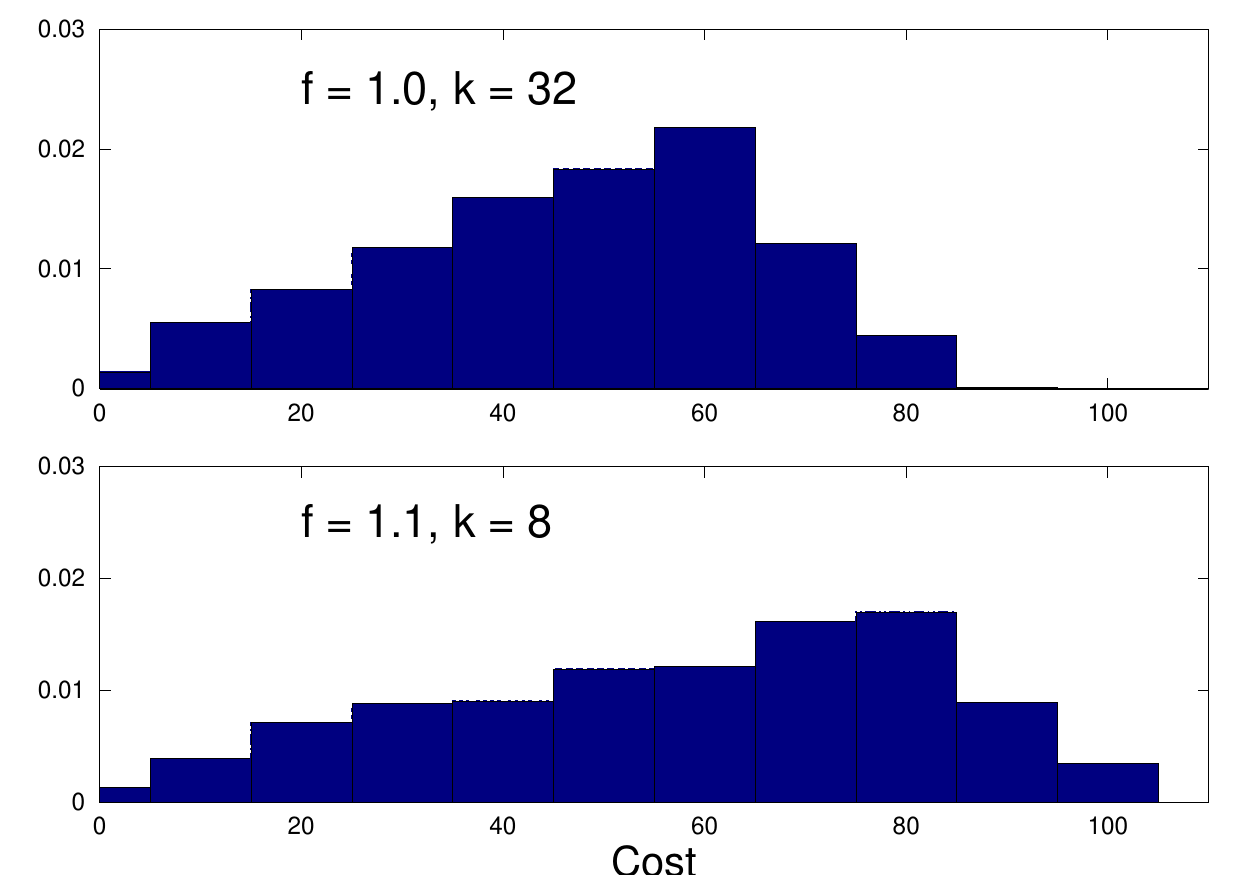}}  
\end{center}
\caption{
Histograms showing distribution of cost for each node to transmit data
to the sink during two simulations.
Upper and lower histogram are for simulations 
with $f = 1, k = 32$ 
and $f = 1.1, k = 8$ (''rapid method'') respectively.
}
\label{fig:hist_cost_e}
\end{figure}

The simulations with $f = 1.1$ resulted in a tree network 
which is not cost optimal to transport of data to the sink 
(Figure \ref{fig:hist_cost_e} shows the distribution of cost). 

\begin{figure}[htp]
\begin{center}
\resizebox{0.8\textwidth}{!}{\includegraphics{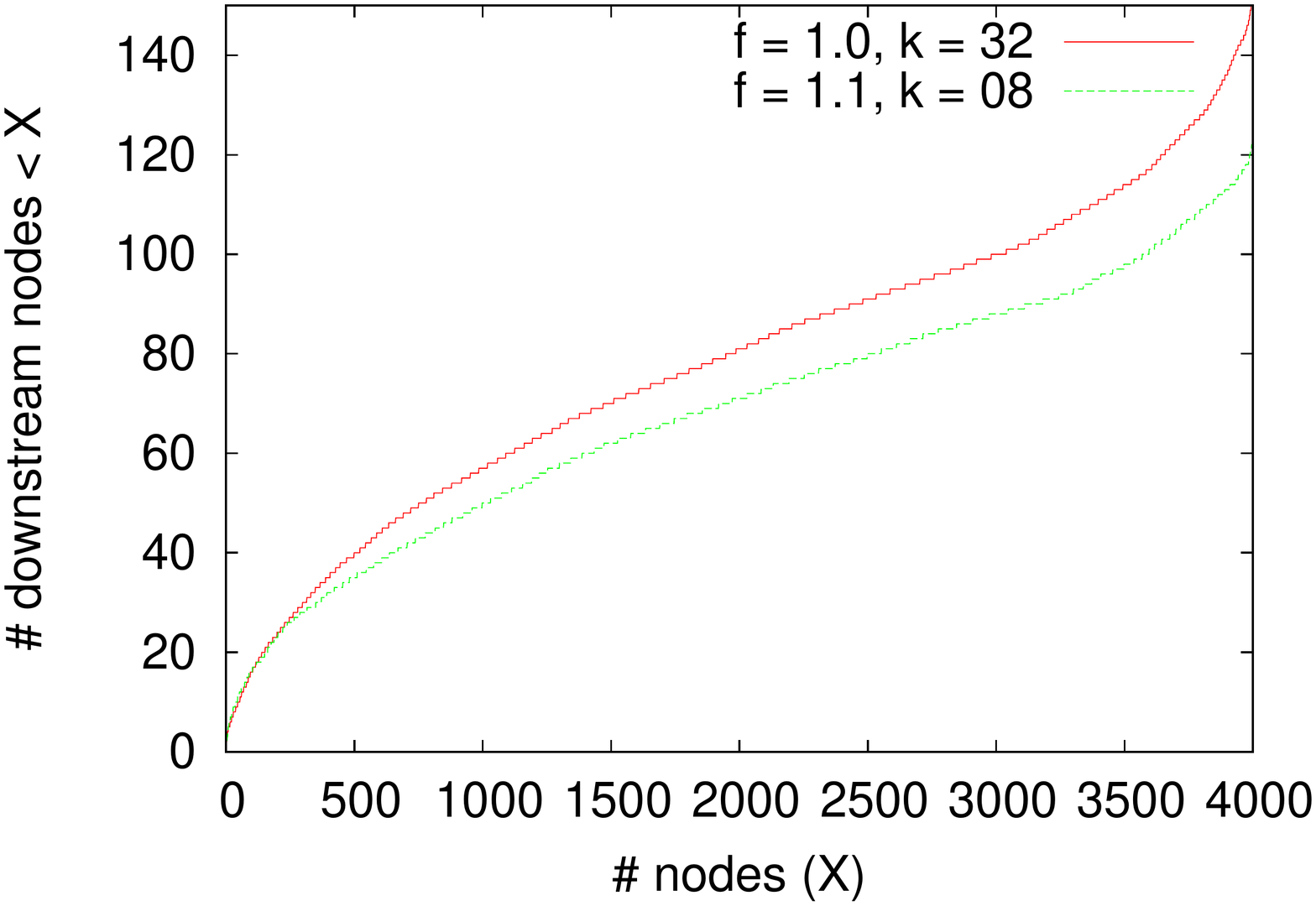}} 
\end{center}
\caption{Distributions of the number of nodes on the route from any node 
downstream to the sink.
One of these graphs results from simulations with maximum number of neighbours equal
to 32 and no relaxation on the condition to send link cost updates $f = 1$.
The other simulation is for $f = 1.1$ and $k = 8$ (''the rapid method'').
}
\label{fig:n_upstream}
\end{figure}

Figure \ref{fig:n_upstream} shows the distribution 
of the number of nodes on its route (downstream) to the sink
for two simulations with respectively $f = 1,\, k = 32$ and  $f = 1.1, \, k = 8$
where $k$ is number of neighbours and $f$ is as above. 
The figure shows that the most rapid method ($f = 1.1,\, k = 8$)
actually gives fewer number of nodes on the route down to the sink
as compared to the more comprehensive search method ($f = 1,\,k=32$).
One may guess the opposite would take place since 
restricting the neighbourhood to the few
nearest neighbours would make more small hops. 

\begin{figure}[htp]
\begin{center}
\resizebox{0.99\textwidth}{!}{\includegraphics{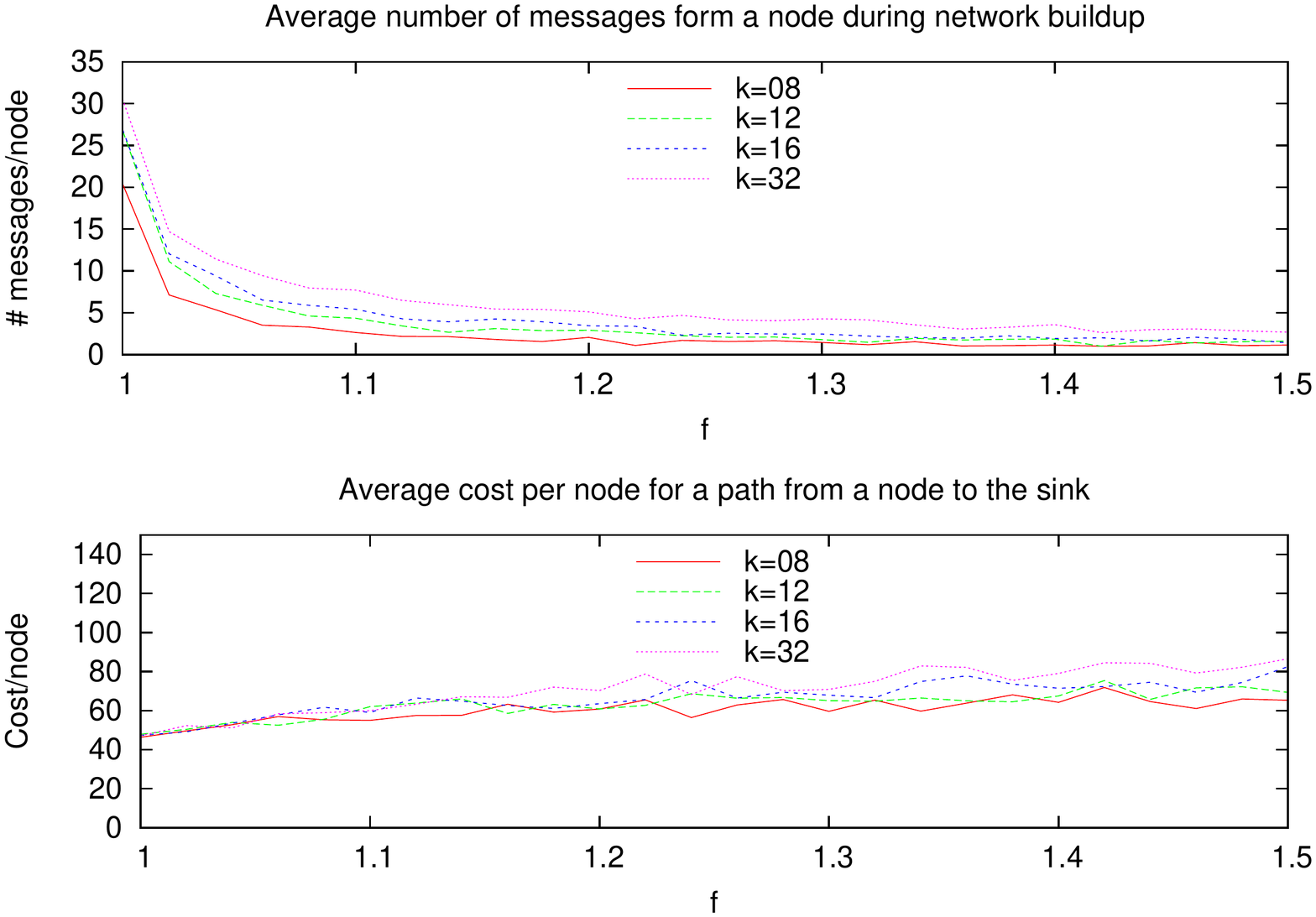}} \\
\end{center}
\caption{Mean number of messages from a node and mean 
         cost per path (from a node to the sink) as a function
         of the threshold $f$  (cf Eq.\ \ref{eq:test1_general} ) 
         and $k = 8, 12, 16, 32$.}
\label{fig:mean_f_k}
\end{figure}

Figure \ref{fig:mean_f_k} shows average values for the 4000 nodes in the present simulations. 
It shows (average) number of messages from a node $X$ and the cost $C(X,S)$  
of the path from the node and to the sink. 
These (node average) values are for the threshold $f$ in the range 1 to 1.5 
and network degree $k = 8, 12, 16,32$.
The (average) number of messages per node here
increases from about 2 for $k = 8$ and $f = 1.2$ to about 30
for $k = 32$ and $f = 1.0$.  
Such a significant change in number of messages suggests a ''regime shift'' in
the network signalling without similar change in cost.

\section{Discussion}
The present sink direction protocol 
has similarities to heat conduction (elliptic problems).
The application of \mbox{Eq.\ \ref{eq:test1}} resembles solving 
a heat equation where the cost estimate $C(X,S)$ at a node $X$
is the temperature which relaxes (adapts) to the neighbourhood.
The application of \mbox{Eq.\ \ref{eq:test1_general}} similarly resembles 
an energy distribution with a quantum (barrier) for small scale
energy movements. 
One may speculate if the Ising model 
can be used to make fast and favourable self-organised
long range correlations in sensor networks and minimise 
network related control traffic.

\begin{figure}[htp]
\begin{center}
\resizebox{0.8\textwidth}{!}{\includegraphics{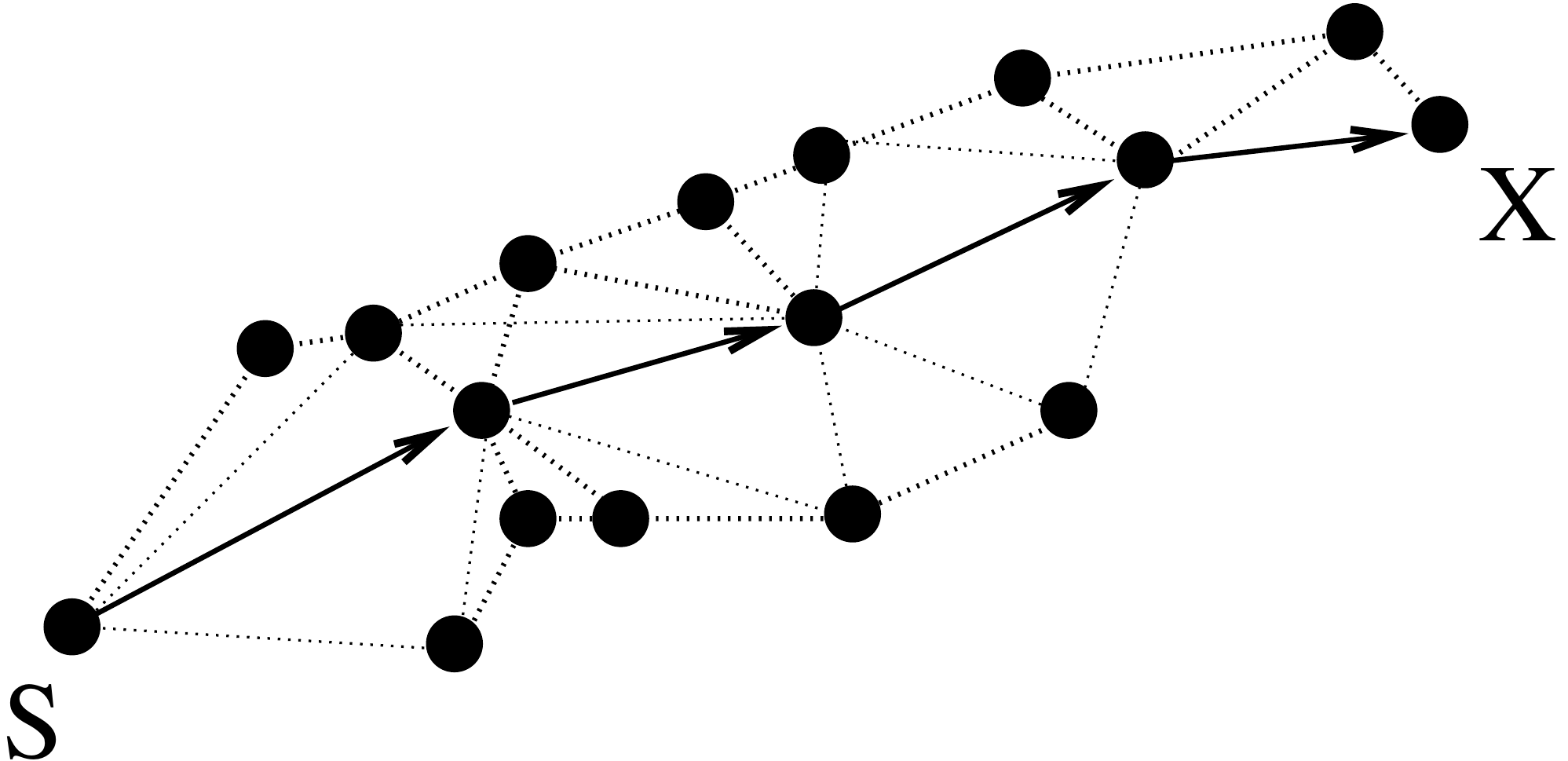}} 
\end{center}
\caption{Possible path from the sink $S$ and to a node $X$.
The solid line illustrates a final (almost optimal) path between $S$ and $X$.
Dotted lines illustrate temporary pointers (edges) during network
organisation which starts gradually from $S$.
Assume a threshold $f > 1$ (cf Eq.\ \ref{eq:test1_general}) for link update 
advertisement and network degree $k$ just large enough
to ensure global network percolation.
The growth in this case tends, for each step, 
more to follow a stable (and close to optimal) route as compared to the situation  
$f = 1$ and larger values for $k$.
This tendency increases as the path approaches $X$
since advertisement from a node at the path to $X$ may affect
possible nodes at the path closer to $X$. 
}
\label{fig:optim1}
\end{figure}

Figure \ref{fig:optim1} illustrates how long range correlation 
can emerge in a situation of a threshold $f > 1$ and network degree $k$
close to the threshold for global network percolation.
For these values of $f$ and $k$,
there is a larger probability that 
the process will more directly (with less updates) 
give the shortest/stable path as compared to otherwise 
(i.e.\ for $f = 1$ and $k$ larger than
necessary for global network percolation).
This tendency will increase for subsequent 
link updates as the tree approaches the node $X$ far from $S$.

This work assumes, for simplicity, no packet loss.
Packet loss may affect convergence,
but it seems intuitively not to affect the main conclusions of this work.
Lost messages can cause delay and redundant transmissions.
 
The sink direction protocol generates pointers forming a tree network 
leading to a data collection unit (sink node). 
Data may in this situation be lost if a node do not perform its task 
to pass on data packets.
Figure \ref{fig:n_upstream} shows the distribution of number of upstream nodes
within the (simulated) network of Figure \ref{fig:map_cost}.
Both of these figures indicate that data from large parts of the network may
not reach the sink node if another node close to it malfunctions.
However, several nodes can normally only by listening detect 
whether a node 
within its neighbourhood 
do not perform its data packet relay function. 
Hence one of the neighbouring nodes may take over its function to relay data. 
Data streams (use of the network) may constantly be used to optimise 
the network (reorganisation of pointers) if data packets contain 
information about (one step) 
transmission cost (given by for example \mbox{Eq.\ \ref{eq:acoustic_cost}}). 
This also gives opportunities for fine tuning of transport ways.
Note that this behaviour is similar to biological systems.

The present protocol may allow nodes to be passive in data relaying
for example by simply not advertising their cost estimates.
These nodes can still take part in data collection as  
(leaf) nodes in the tree network (but not initially relay data).

\bibliographystyle{elsart-num}


\end{document}